\documentclass[prl,twocolumn,showpacs,showkeys]{revtex4}
\usepackage{graphicx}
\usepackage{dcolumn}
\usepackage{bm}

\begin{document}

\title{Tunable high-energy ion source via oblique laser pulse incidence on a
double-layer target}
\author{T.~Morita}
\author{T.~Zh.~Esirkepov}
\author{S.~V.~Bulanov}
\author{J.~Koga}
\author{M.~Yamagiwa}
\affiliation{Japan Atomic Energy Agency, 8-1 Umemidai, Kizugawa, Kyoto 619-0215, Japan}

\begin{abstract}
The laser-driven acceleration of high quality proton beams from a
double-layer target, comprised of a high-Z ion layer and a thin disk of
hydrogen, is investigated with three-dimensional particle-in-cell
simulations in the case of oblique incidence of a laser pulse. It is shown
that the proton beam energy reaches its maximum at a certain incidence angle
of the laser pulse, where it can be much greater than the energy at normal
incidence. The proton beam propagates at some angle with respect to the
target surface normal, as determined by the proton energy and the incidence
angle.
\end{abstract}

\maketitle

The method of charged particle acceleration by using laser light is very
attractive, since the acceleration rate is much higher and the facility size
can be substantially smaller than standard accelerators. Ion acceleration
during the high intensity electromagnetic wave interaction with plasmas was
proposed more than 50 years ago \cite{50ies}. Currently, ion acceleration
experiments using high power lasers close to petawatt levels are going on
all over the world \cite{SKN}. Laser driven fast ions are considered in
regard to applications ranging from hadron therapy \cite{SBK}, fast ignition
of thermonuclear targets \cite{ROT}, production of PET sources \cite{SPN},
conversion of radioactive waste \cite{KWD}, a laser-driven heavy ion
collider \cite{ESI1}, injectors for standard accelerators \cite{INJ}, and
proton radiography \cite{PrIm} to proton dump facilities for neutrino
oscillation studies \cite{SVB} (see Refs. \cite{REVS} and literature quoted
therein).

The typical energy spectrum of laser accelerated particles from unoptimized
targets is thermal-like, with a cut-off at a maximum energy. On the other
hand, almost all the above mentioned applications require high quality
proton beams, i.e. beams with sufficiently small energy spread $\Delta  
\mathcal{E}/ \mathcal{E}\ll 1$. As suggested in Ref. \cite{DL}, such a
required beam of laser accelerated ions can be obtained using a double-layer
target, which consists of high-Z atoms and a thin coating of low-Z atoms.
Extensive computer simulations of this target were performed in Refs. \cite%
{ESI} and \cite{EYT}, where multi-parametric particle-in-cell (PIC)
simulations were used to optimize the laser-driven proton acceleration by
choosing appropriate laser and target conditions. The feasibility of the
double-layer target scheme was demonstrated experimentally with
microstructured targets in Ref. \cite{SCH}. The effects of target shaping on
the laser-driven ion acceleration were also reported in Refs. \cite{SON}.
Previously, the double-layer target scheme for high and controllable quality
ion acceleration has been mostly studied in the configuration of normal
incidence of the laser pulse on the target. However, the case of oblique
incidence provides an additional parameter for manipulation of the fast ion
energy, the emittance, energy spectrum and the proton beam propagation
direction. As is well known, the energy transfer from a p-polarized
obliquely incidence electromagnetic wave to the electron energy via the
so-called "vacuum heating" mechanism \cite{BRU} depends on the incidence
angle and is substantially higher than for normal incidence. A stronger
electron heating results in a stronger electric field generation due to the
electric charge separation effect, which in turn leads to more efficient ion
acceleration.

In this Letter, we study the dependence of the ion beam energy and quality
on the laser incidence angle. We use an idealized model, in which a Gaussian
p-polarized laser pulse is incident on a double-layer target of
collisionless plasmas. The simulations are performed with a
three-dimensional massively parallel electro magnetic
code, based on the PIC method \cite{CBL}. The number of grid cells is equal
to $2800\times 720\times 720$ \ along the $x$, $y$, and $z$ axes. The total
number of quasi-particles equals $7\times 10^{7}$.
The size of the
simulation box is $100\lambda \times 25.5\lambda \times 25.5\lambda $, where 
$\lambda$ is the laser wavelength. The boundary conditions for the particles
and for the fields are periodic in the transverse ($y$,$z$) direction and
absorbing at the boundaries of the computation box along the $x$ axis. Here
the laser wavelength determines the transformation from dimensionless to
dimensional quantities and vice versa. Below, the dimensional quantities are
given for $\lambda = 0.8\mu $m; the spatial coordinates are normalized to $%
\lambda $ and the time is measured in the laser period, $2\pi/\omega$.

\begin{figure}[tbp]
\includegraphics[width=8.6cm,bb=0 0 2450 2480]{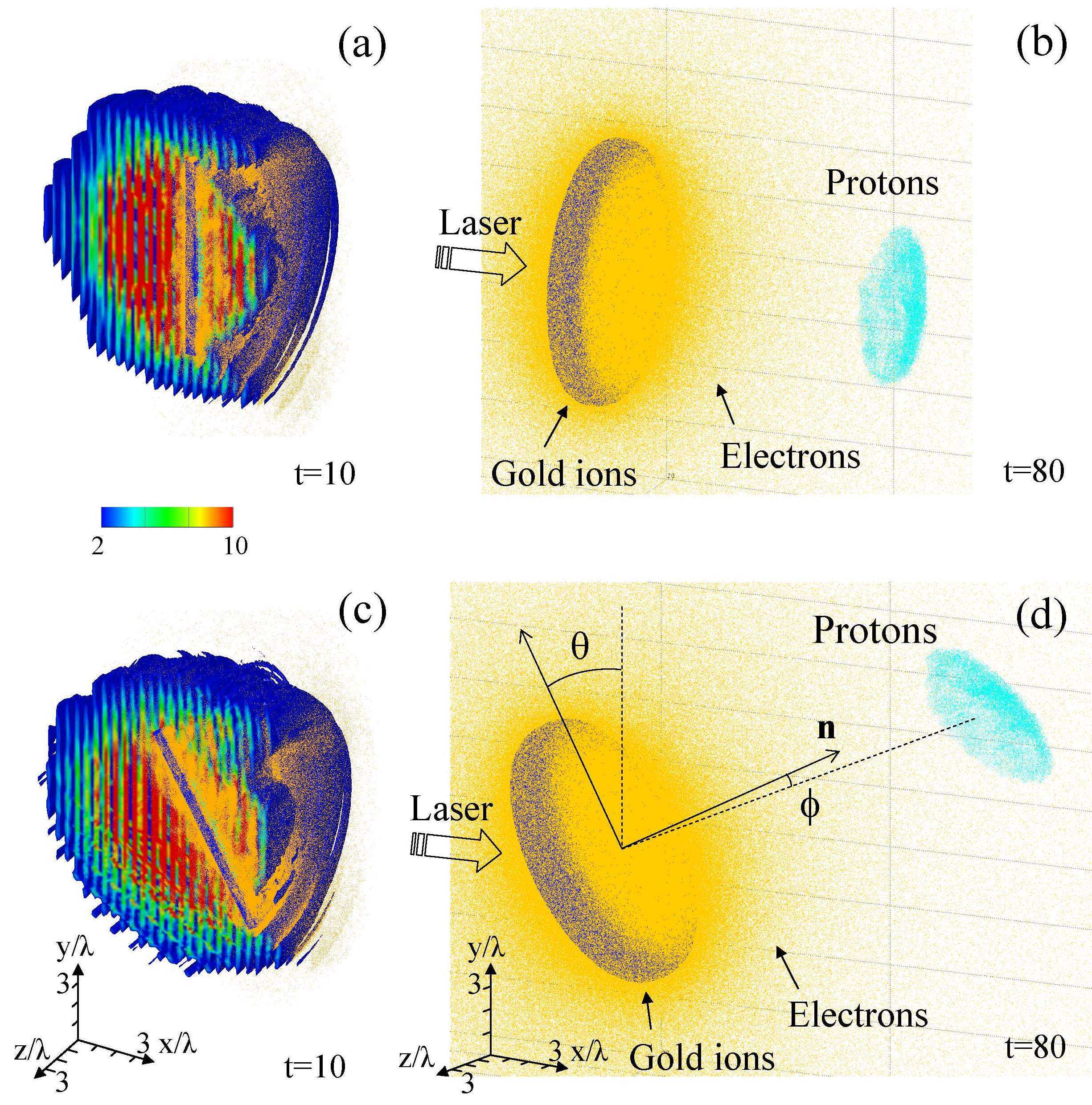}
\caption{Proton acceleration driven by the laser pulse with normal (a,b) and
oblique (c,d) incidence. (a,c): Electric field magnitude (isosurface for
value $a=2$) at t=10; half of the box is removed to reveal the internal
structure. (b,d): Distribution of gold ions (blue), electrons (yellow), and
protons (light blue) at t=80. The laser pulse incidence angle in (c,d) is $%
\protect\theta=30^{ \mathrm{o}}$.}
\label{fig:A}
\end{figure}

The Gaussian laser pulse with the dimensionless amplitude $%
a=eE_{0}/m_{e}\omega c=30$, which corresponds to the laser peak intensity $%
2\times 10^{21}W/cm^{2}$, is $8\lambda $ long in the propagation direction
and it is focused to a spot with size $6\lambda $ (FWHM). The oblique
incidence of the laser pulse is realized by tilting the target around the $%
z- $axis, Fig. \ref{fig:A}(d), while the laser pulse propagates along the $x-
$ axis. Both layers of the double-layer target are shaped as disks. The
first, gold, layer has a diameter $10\lambda $ and thickness $0.5\lambda $.
The second, hydrogen, layer is narrower and thinner; its diameter is $%
5\lambda $ and thickness is $0.03\lambda $. The electron density inside the
gold layer is $n_e = 1.6\times 10^{22}$cm$^{-3}$; inside the proton layer it
is $n_e = 5\times 10^{20}$cm$^{-3}$.

Figure \ref{fig:A} shows the proton beam acceleration for two cases of
normal and oblique incidence. For the present simulation parameters, the
target is partially transparent for the electromagnetic wave (due to the
relativistic transparency effect), which, according to Refs. \cite{ESI} and 
\cite{EYT}, corresponds to an optimal condition of the ion acceleration. In
the case of oblique incidence, a substantially larger portion of the
electrons is blown off the target due to the effect of "vacuum heating."

\begin{figure}[tbp]
\includegraphics[width=8.6cm,bb=0 0 1581 1437]{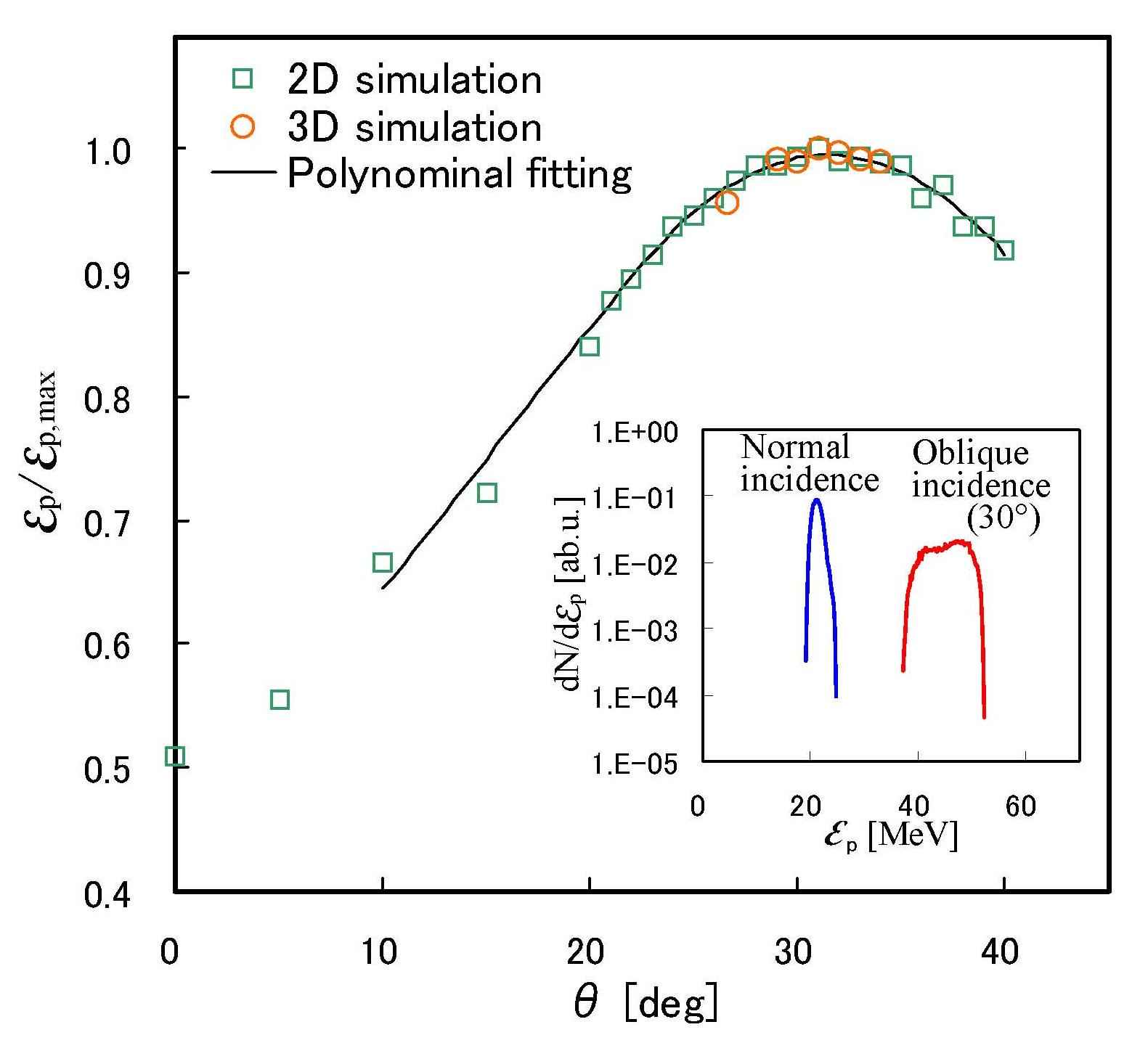}
\caption{Proton energy, normalized by the maximum, vs the laser pulse
incidence angle. In the inset: proton energy spectra at t=80, as obtained in
3D PIC simulation.}
\label{fig:B}
\end{figure}

In order to examine the dependence of the energy achieved by the protons on
the incidence angle, $\theta $, we performed extensive 2D and 3D PIC
simulations. Fig. \ref{fig:B} shows the proton energy as a function
of the laser pulse incidence angle, normalized by the maximum achievable
energy. We see that the maximum proton energy is reached at the incidence
angle approximately equal to 30$^{ \mathrm{o}}$, for which case we undertook
3D PIC simulation shown in Fig. \ref{fig:A}(c,d). The proton energy value is
approximately doubled at $\theta $=30$^{ \mathrm{o}}$ compared to the case
of normal incidence, $\theta $=0$^{ \mathrm{o}}$. This also can be seen in
the inset in Fig. \ref{fig:B} showing the proton energy spectra at t=80.
While the proton energy increased from $\sim $20 MeV at normal incidence to $%
\sim $45 MeV at oblique incidence with an optimal angle, the energy spread
is also increased from 10\% to 23\%.

If we invoke the above mentioned "vacuum heating" mechanism for the laser
pulse energy transformation into the electron component energy, and then,
via acceleration in the electric field of the charge separation, into the
energy of fast protons, we find that the laser pulse energy deposited to the
target \ and the efficiency of pushing the electrons out from the target
depend on the incidence angle $\theta $. The counterplay between these two
effects leads to the formation of the fast proton energy maximum at a
certain incidence angle. Under the conditions of our simulations this angle
is approximately equal to 30$^{ \mathrm{o}}$. The analysis of the time
evolution of the proton energy spectrum shows that both the average energy
and energy spread increase with time. The energy spread appears to grow in
the oblique incidence case, which can be accounted for by the asymmetry of
the quasistatic electric field along the target surface at oblique
incidence. We note that, according to Ref. \cite{DL}, the energy spread can
be decreased by reducing the thickness and transverse size of the low-Z ion
(proton) layer.

\begin{figure}[tbp]
\includegraphics[width=8.6cm,bb=0 0 1581 1431]{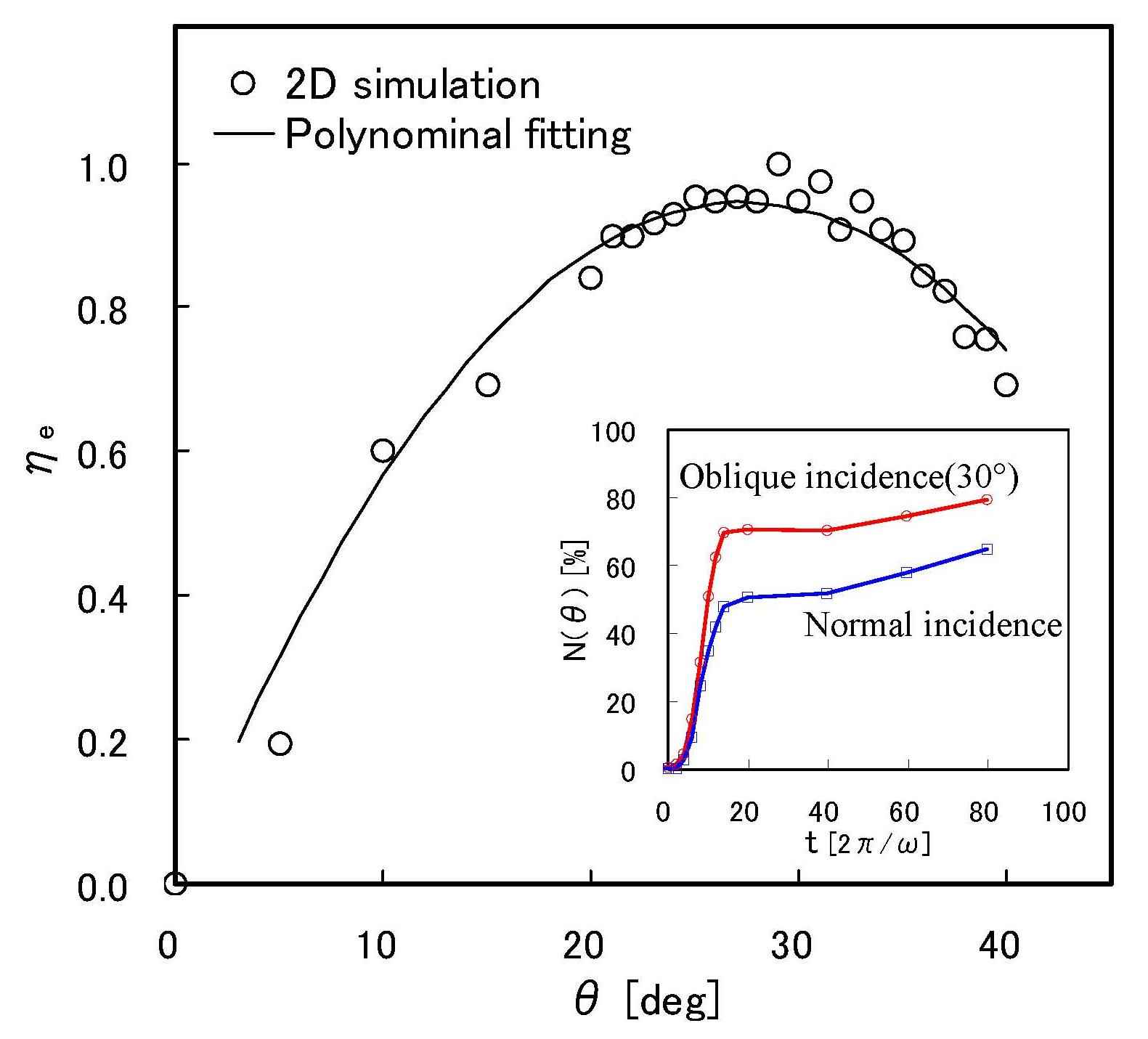}
\caption{Escaping electron ratio $\protect\eta_e$ vs the laser pulse
incidence angle at t=20. In the inset: normalized number of electrons swept
off the target vs time, as obtained in 3D PIC simulation. }
\label{fig:C}
\end{figure}

\begin{figure}[tbp]
\includegraphics[width=6.0cm,bb=0 0 1566 1419]{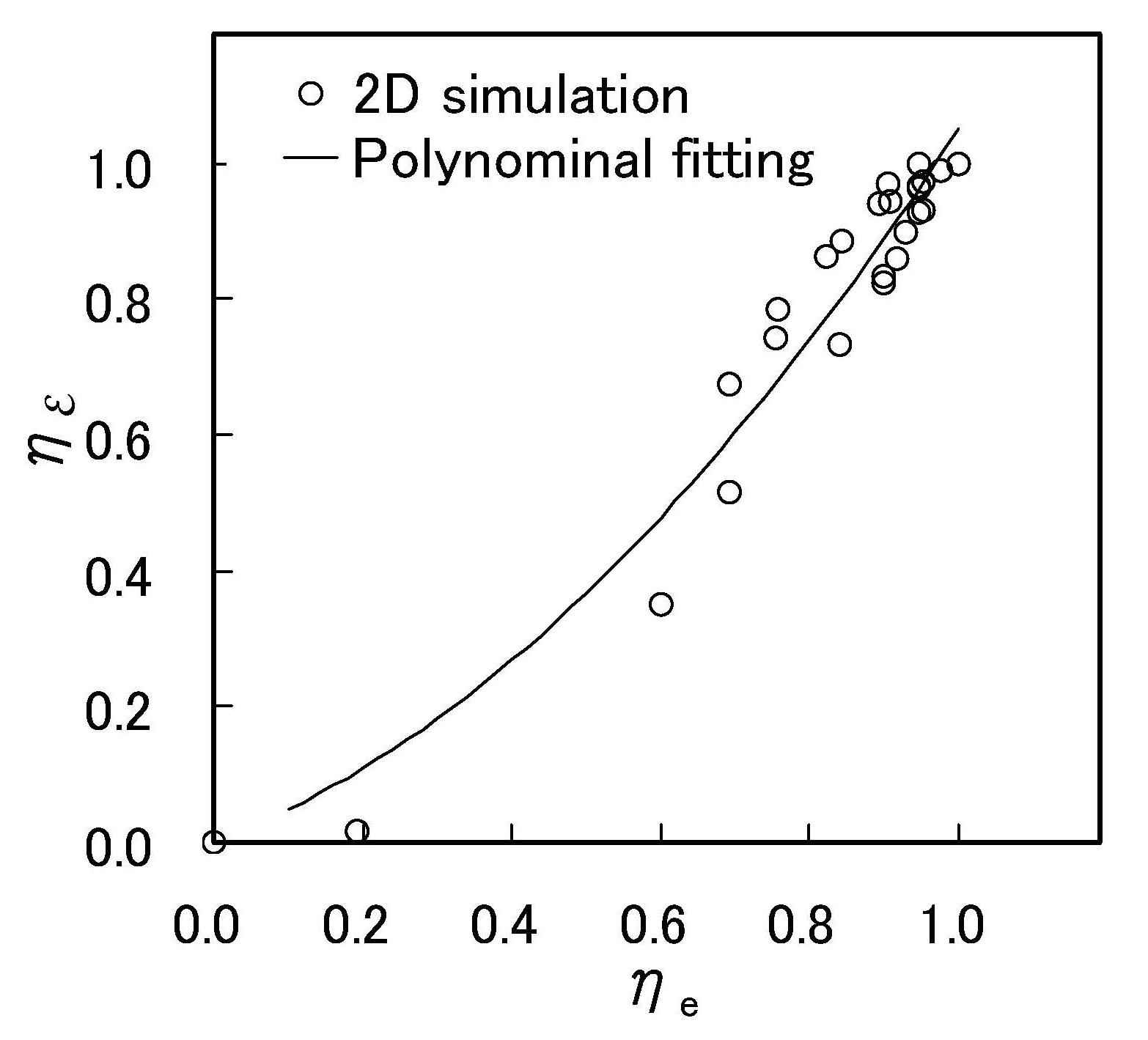}
\caption{Proton energy ratio vs escaping electron ratio at t=20. }
\label{fig:D}
\end{figure}

Since the proton acceleration occurs due to the electric field generation
through the electron ejection from the target, it is important to analyze
the dependence of the number of the electrons leaving the target under the
laser radiation action on the incidence angle. Figure \ref{fig:C} shows the
ratio of electrons pushed out from the target versus the incidence angle at
time t=20. The dependence at t=20 is chosen because at this time the number
of ejected electrons saturates as seen in the inset in Fig. \ref{fig:C}. The
ratio of ejected electrons as a function of the incidence angle $\theta $ is
defined by the expression $\eta_{e}=(N_{e}(\theta
)-N_{e,min})/(N_{e,max}-N_{e,min})$, where $N_{e}(\theta )$ is the number of
ejected electrons, $N_{e,max}=\max\{N_{e}(\theta )\}$ and $%
N_{e,min}=\min\{N_{e}(\theta )\}$ are the maximum and minimum values,
respectively. We see that this ratio reaches its maximum at the incidence
angle about 30$^{ \mathrm{o}}$, which evidences a strong correlation between
the dependences of the number of ejected electrons (Fig. \ref{fig:C}) and
the proton energy (Fig. \ref{fig:B}) on the laser pulse incidence angle. We
see from Fig. \ref{fig:C} that the efficiency of the electron ejection is
substantially higher for $\theta $=30$^{ \mathrm{o}}$ than in the case of
normal incidence. As a consequence, a stronger electric field is produced
and the protons are accelerated to higher energy. This is also illustrated
in Fig. \ref{fig:D}, by the correlation between the average proton energy
and the ejected electron number. Here the proton energy ratio at $\theta $
is defined by the expression $\eta_{ \mathcal{E}}=( \mathcal{E}_{p}(\theta
)- \mathcal{E}_{p,min})/( \mathcal{E}_{p,max}- \mathcal{E}_{p,min})$, where $%
\mathcal{E}_{p}(\theta )$ is the average proton energy, $\mathcal{E}%
_{p,max}=\max\{ \mathcal{E}_{p}(\theta )\}$ and $\mathcal{E}_{p,min}=\min\{ 
\mathcal{E}_{p}(\theta )\}$ are the maximum and minimum values,
respectively. We see that the fast proton energy is roughly proportional to
the number of ejected electrons.

As seen from Fig. \ref{fig:A}, the accelerated proton bunch retains the form
of the disk. In the case of the normal incidence, the proton disk moves in
the direction normal to the target surface while its surface remains
parallel to the first (high-Z ion) layer. In the case of oblique incidence,
the proton disk motion direction deflects from the target normal by a
noticeable angle, $\phi $, while the disk surface is tilted with respect to
the high-Z ion layer, Fig. \ref{fig:A}(d). We note that a deflection of the
accelerated proton bunch has been observed in the experiments on the laser -
solid target interaction \cite{Zepf}. Under the conditions of our
simulations, which correspond to the relatively higher laser intensity, the
deflection and tilt can be explained by relativistic effects. As is known,
the Lorentz transformation to the frame of reference moving with the
velocity $V=-c\sin \theta $ along the target surface, i.e. the gamma factor
is given by $\gamma =1/\sqrt{1-V^{2}/c^{2}}=1/\cos \theta $, changes the
configuration of the wave--target interaction from oblique to normal
incidence \cite{BOUR}, so that the wave frequency and wave vector of the
incident electromagnetic wave become $\omega ^{\prime }=\omega \cos \theta $
and $\mathbf{k}^{\prime }=k\cos \theta \mathbf{e}_{||}$, where $\mathbf{e}%
_{||}$ is the unit vector along $\mathbf{k}^{\prime }$. In this new boosted
reference frame, the protons have a transverse component of momentum equal
to $p_{\bot }^{\prime }=-m_{p}c\tan \theta $, where $m_{p}$ is the proton
mass. As a result of the acceleration in the electric charge separation
field, the protons acquire the longitudinal momentum $p_{||}^{\prime }$ and
their energy becomes equal to $\mathcal{E}_{p}^{\prime }=\sqrt{%
m_{p}^{2}c^{4}+p_{||}^{\prime 2}c^{2}+p_{\bot }^{\prime 2}c^{2}}$.
Performing the Lorentz transformation back to the laboratory reference
frame, we obtain that $p_{||}=p_{||}^{\prime }$, $p_{\bot }=\gamma
(p_{||}^{\prime }-\mathcal{E}_{p}^{\prime }V/c^{2})$ and the deflection
angle $\phi =\arctan (p_{\bot }/p_{||})$ is equal to 
\begin{equation}
\phi =\arctan \left[ \frac{m_{p}c\tan \theta }{p_{||}\cos \theta }\left( 
\sqrt{1+\left( \frac{p_{||}\cos \theta }{m_{p}c}\right) ^{2}}-1\right) %
\right]   \label{phi}
\end{equation}%
In the limit $p_{||}/m_{p}c\ll 1$, which corresponds to the parameters of
our simulations, this expression yields $\phi \approx \sqrt{\mathcal{E}%
_{p}/2m_{p}c^{2}}\sin \theta $. For $p_{||}/m_{p}c\gg 1$, the angle $\phi $
becomes $\phi \approx \theta $, i.e. the protons are accelerated almost
along the laser pulse propagation direction.

In order to account for the proton disk tilting, we note that the obliquely
incident laser pulse front propagates along the target surface with a
superluminal velocity $V_{F}=c/\sin \theta $. The time delay between moments
when different disk elements separated by the distance $\Delta l$ start to
move is equal to $\Delta t=\Delta l\sin \theta /c$. The displacement of the
proton layer elements in the direction of the target normal can be estimated
as $\Delta \xi _{||}=p_{||}\Delta t/m_{p}$. This gives the angle of the
proton disc tilt, $\chi =\arctan (\Delta \xi _{||}/\Delta l)$, i.e. $\chi
\approx \sqrt{2 \mathcal{E}_{p}/m_{p}c^{2}}\sin \theta $. This effect is
seen in the simulations at an early time of the proton acceleration. Later,
higher dimensional effects come into play and the tilting angle changes.

\begin{figure}[tbp]
\includegraphics[width=6.0cm,bb=0 0 1578 1446]{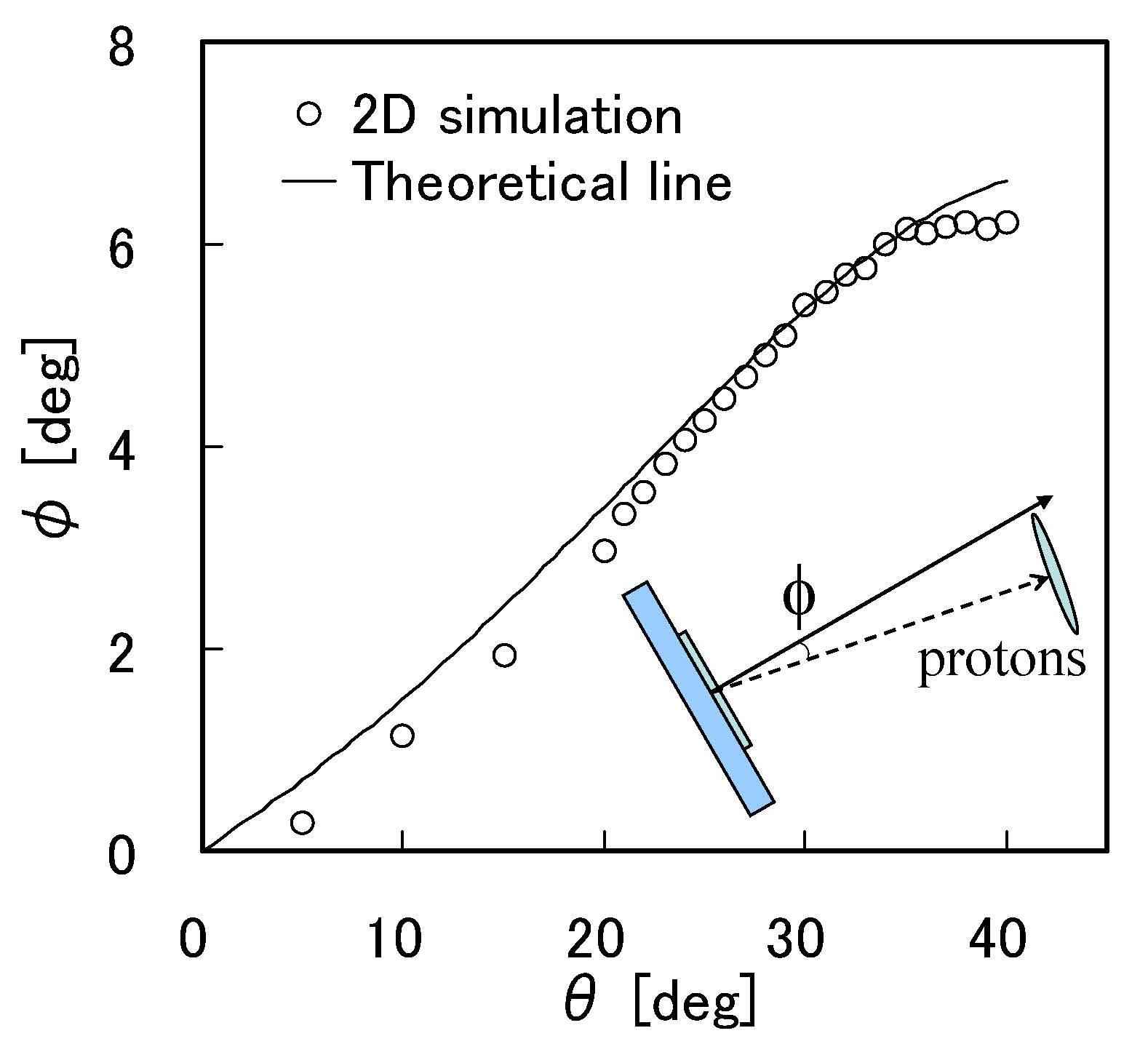}
\caption{Deflection angle of the proton bunch vs the laser pulse incidence
angle at t=80. The solid curve is given from Eq. (\protect\ref{phi}). }
\label{fig:E}
\end{figure}

The dependence of the angle of the deflection of the proton motion from the
target normal, $\phi $, on the laser pulse incidence angle, $\theta $, is
seen in Fig. \ref{fig:E}. Here the deflection angle, $\phi $, is defined as
the angle between the normal to the target surface placed at the target
center and the average radius-vector from the target center to the proton
layer, which equals $\Sigma _{i=1}^{N_{p}}\mathbf{x}_{pi}/N_{p}$, where $%
\mathbf{x}_{pi}$ is the radius-vector of the $i$-th proton out of $N_{p}$
protons in the accelerated beam. As seen from Eq. (\ref{phi}), in the limit
of a small incidence angle, $\theta $, the proton bunch is accelerated
almost along the direction normal to the target with $\phi $ being a linear
function of $\theta $. When the incidence angle, $\theta $, increases, the
growth of the deflection angle, $\phi $, saturates, in agreement with the
results of simulations, shown in Fig. \ref{fig:B}.

In conclusion, we have found that the proton acceleration during laser pulse
interaction with double-layer targets is more efficient for oblique
incidence than for normal incidence. It is shown that the proton beam energy
reaches its maximum at a certain incidence angle of the laser pulse, where
it can be much greater than the energy at normal incidence. The proton beam
propagates at some angle with respect to the target surface normal, as
determined by the proton energy and the incidence angle. In the limit of
nonrelativistic proton energy, the deflection angle is relatively small.
However, its value $\approx 6^{ \mathrm{o}}$ (see Fig. \ref{fig:E}) results
in the proton beam deflection of the order of 10 cm after the protons have
propagated over 1 m distance.
Therefore this angle should be taken into account in the real facility design
to make proper mutual location of the proton source and the target.
This provides a way to control the proton
energy and the direction of the proton beam propagation by adjusting the
incidence angle of the laser pulse. Such flexibility and speed in
controlling the proton directionality and energy will be invaluable for
conformal cancer therapy treatments such as pixel scanning using heavy ions 
\cite{HBS}.

Two of authors (S.V.Bulanov,T.Zh.Esirkepov) thank O.Willi for useful discussions.
This work was in part supported by JST/CREST and MEXT. The computation was
done with the ES40 at JAEA Kansai and ALTIX 3700 at JAEA Tokai.

\end{document}